\documentclass[prc,preprint]{revtex4}

\usepackage{graphicx}

\begin{document}

\title{{\bf Radial deformation of the Kerr spacetime}}
\author{T. Ghasabi\footnote{E-mail:
ghasabitm@yahoo.com} and N. Riazi\footnote{E-mail:
n\_riazi@sbu.ac.ir} } \affiliation{Physics Department, Shiraz
University, Shiraz 71454, Iran} \affiliation{Physics Department,
Shahid Beheshti University, Evin, Tehran 19839, Iran}



\begin{abstract}
Since the Kerr metric is an idealized vacuum solution, the
spacetime around accreting rotating black holes is certainly a
non-vacuum solution and thus deviates from the Kerr metric.  In
the absence of an exact interior Kerr solution, we propose a
radial deformation of the Kerr metric  which leads to a non-vacuum
geometry and study the spacetime as a function of the deformation
parameter. The status of energy conditions is discussed in detail.
It is shown that the resulting spacetime leads to the appearance
of anisotropic wormholes in certain cases.
\end{abstract}

\maketitle
\section{INTRODUCTION}
Black holes are perhaps the most strange and fascinating objects
known to exist in the universe that are predicted by Einstein's
theory of gravity. Though these objects made their first
appearance in the famous exact spherically symmetric solution
found by Karl Schwarzschild \cite{1}, the concept of black hole
has been crystallized by physicists for many years \cite{3p}. For
a long time, a part of the physics community was rather sceptical
about the actual existence of black holes, but the situation has
changed in recent years, notably because of the wide variety of
phenomena of astronomical observations: from X-ray binary systems
to active galactic nuclei (including our home, the Milky Way)
\cite{3}.

The problem of finding a  vacuum gravitational field which
describes a stationary rotating black hole was solved by Kerr
\cite{Kerr} and his solution remains of considerable interest in
general relativity. In astrophysics, it is very important to have
a metric for the interior of a rotating star, which should match
to the solution of Kerr, since we usually take the solution of
Kerr as an exterior solution. Many people tried to find interior
solutions but what they found had in general some problem. For
example, the matching was approximate \cite{5rot, 7rot}. It was
found that regular disks as sources of the Kerr metric give
energy-momentum tensors which do not satisfy the dominant energy
conditions \cite{8rot}. The question if a perfect fluid can be the
source of the metric of Kerr is an open one \cite{9rot, 10rot}.

There are strong evidences that black holes and wormholes are
closely related. A new frame for the unity of the black hole with
the wormhole's dynamics is presented by Hayward \cite{79me}. Black
holes are described with an outer null trapped surface and
wormholes with an outer timelike trapped surface that the incoming
light beam from the surface will start to diverge \cite{80me}.
Wormholes are tunnels in the geometry of space and time that
connect two separate and distinct regions of spacetimes. Although
such objects were long known to be solutions of Einstein equation,
a renaissance in the study of wormholes took place during 80?s
motivated by the possibility of quick interstellar travel
\cite{MT}.

Currently, there exist some activities in the field of wormhole
physics following, particularly, the seminal works of Morris,
Thorne and Yurtsever \cite{13hen}. Morris and Thorne assumed that
their traversable wormholes were time independent, non-rotating,
and spherically symmetric bridges between two universes. The
manifold of interest is thus a static spherically symmetric
spacetime possessing two asymptotically flat regions. These kinds
of wormholes could be threaded both by quantum and classical
matter fields that violate certain energy in the vicinity of the
throat known as exotic matter. On general grounds, it has recently
been shown that the amount of exotic matter needed at the wormhole
throat can be made arbitrarily small, thereby facilitating an
easier construction of wormholes \cite{14hen}.

Previously, Damour and Solodukhin investigated a radial
deformation of Schwarzschild spacetime, showing that the resulting
wormholes can mimic many observational features of black holes
\cite{Damour}. A radial deformation of the Reissner-Nordstr"m
metric which leads to the appearance of charged, traversable
wormholes was investigated by Mehdizadeh, Ebrahimi and Riazi
\cite{me}. Lemos et al. \cite{lemos} studied  extremal
"$\epsilon$-wormholes on the threshold of the formation of an
event horizon, quasi-black holes, and wormholes on the basis of
quasi-black holes. They investigated whether or not the resulting
objects remain regular in the near-horizon limit. They showed that
the requirement of full regularity, i.e., finite curvature and
absence of naked behavior, up to an arbitrary neighborhood of the
gravitational radius of the object implies ruling  out potential
mimickers in most of the cases.

In this article, we apply a radial deformation to the Kerr metric
and the corresponding energy- momentum tensor treading this metric
is calculated which gives the required type of matter distribution
and energy flux for this deformation. Energy conditions and the
question whether this new structure is a deformed black hole or it
is a wormhole, is investigated. In other words, we present a
family of rotating anisotropic asymptotically flat fluid
solutions, which depend on three parameters including the
parameters of mass and angular momentum. For certain values of
these parameters, although $\rho>0$, the weak (and therefore null
and strong) energy conditions are violated in the whole physical
spacetime. All solutions are singular on a ring lying in the
equatorial plane (like the ring singularity of Kerr solution),
while the deformed metric has one more singularity.

\section{deforming the kerr spacetime}
Our motivation for such a deformation is the following: the
presence of a rotating fluid leads to a deviation from the vacuum
Kerr metric. Since it has been proved difficult (or perhaps
impossible) to construct an exact interior Kerr solution, we apply
a radial deformation to the Kerr metric which is supposed to be
caused by the presence of the rotating fluid. In other words,
since in the physical world rotating black holes are not isolated
and accrete matter, the metric will be different from the vacuum
of Kerr metric. Since the investigation of the general form of the
deformed metric is difficult, we study a special kind of radial
deformation.

We consider a metric which in Boyer-Lindquist coordinates
\cite{11} has the form

\begin{eqnarray} \label{eq1}
\nonumber ds^{2}&=&-(1-\frac{2mr}{r^{2}+a^{2}\cos^{2}\theta})dt^{2}+(\frac{r^{2}+a^{2}\cos^{2}\theta}{r^{2}+2r^{2}\varepsilon+r^{2}\varepsilon^{2}-2mr-2mr\varepsilon+a^{2}})dr^{2}\\
&+&(r^{2}+a^{2}\cos^{2}\theta)d\theta^2+
\sin^{2}\theta(r^{2}+a^{2}+\frac{2mra^{2}\sin^{2}\theta}{r^{2}+a^{2}\cos^{2}\theta})d\varphi^{2}-\frac{4mra\sin^{2}\theta}{r^{2}+a^{2}\cos^{2}\theta}dtd\varphi
\end{eqnarray}
where $\rho^{2}=r^{2}+a^{2}\cos^{2}(\theta)$. Here, $m$ and $a$
denote the mass and the rotation parameters, respectively. This
metric differs from the standard Kerr metric \cite{2p, 3p, 4p}
only through the presence of the dimensionless parameter $
\varepsilon $, where $\varepsilon=0$ corresponds to the Kerr
spacetime (endowed with a horizon). By contrast, for
$\varepsilon\neq 0$ the structure of the spacetime is dramatically
different: there is no event horizon, instead there is a throat
that joins two anisotropic and asymptotically flat regions.
This spacetime is an example of a Lorentzian wormhole \cite{10magh}.

The asymptotic expansion of the deformed metric (\ref{eq1})
against r is
\begin{equation}
ds^{2}=-dt^{2}+ \frac{1}{1+ 2\varepsilon
+\varepsilon^{2}}dr^{2}+r^{2}d\theta^{2}+r^{2}\sin^{2}\theta
d\varphi^{2}.
\end{equation}
this relation imply that the deformed metric (\ref{eq1}) is
asymptotically flat for all values of $\varepsilon$. In principle, we must used tetrad component or curvature invariants to discuss the behavior of deformed metric at infinity. The relevant components ( see \cite{3p}) in this case is (according to notations of MTW)
\begin{equation}
F=R_{\theta\varphi}{}^{\theta\varphi}
\end{equation}
and we have
\begin{equation}
R_{\theta\varphi}{}^{\theta\varphi} \propto \frac{1}{r^{2}} \longrightarrow 0 \hspace{.5cm} if \hspace{.5cm} r \longrightarrow \infty
\end{equation}
therefor we conclude that the spacetime is asymptotically flat for all values of $\varepsilon$.

The Kretschmann scalar ($R_{\mu \nu \lambda \kappa}R^{\mu \nu
\lambda \kappa}$) of the deformed metric (\ref{eq1}) is
\begin{equation}
K =
\frac{F(r,\varepsilon,m,a)}{(r^{2}+a^{2}\cos^{2}\theta)^{6}(r^{2}+a^{2}-2mr)^{4}}
,
\end{equation}
where $F$ is a lengthy expression not useful to be reproduced
here (we note that the zeros of F can't compensate the zeros of the denominator. ). The deformed metric (\ref{eq1}) has two intrinsic
singularities which correspond to
\begin{equation}
\rho^{2}=r^{2}+a^{2}\cos^{2}\theta=0
\end{equation}
and
\begin{equation}
\Delta=r^{2}-2mr+a^{2}=0
\end{equation}
these occur when
\begin{equation} \label{eq4}
x^{2}+y^{2}=0 ,\hspace{0.5cm}  z=0
\hspace{0.5cm}(i.e.\hspace{0.5cm} x=y=z=0)
\end{equation}
and
\begin{equation} \label{eq5}
r^{\pm}=m \pm \sqrt{m^{2}-a^{2}}
\end{equation}
 Therefore from (\ref{eq4}) we find that all solutions are
singular on a ring lying in the equatorial plane (the ring
singularity of Kerr solution), while from (\ref{eq5}) we find that
those have two more singularities which corresponds to equation
(\ref{eq5}).

 From the deformed metric (\ref{eq1}), the equation
\begin{equation}
g^{rr}=0
\end{equation}
has two solutions
\begin{equation}
r_{\varepsilon}^{\pm}=\frac{1}{1+\varepsilon}(m \pm
\sqrt{m^{2}-a^{2}}) ,
\end{equation}
while the two solutions of $ g_{tt}=0 $ are
\begin{equation}
r^{\pm}=m \pm \sqrt{m^{2}-a^{2} \cos^{2}\theta} \hspace{.2cm} .
\end{equation}

In order to have real roots, we should have
\begin{equation}
m > |a|
\end{equation}
a condition already known for the Kerr black holes ($m<|a|$
corresponds to naked singularities in the Kerr metric which are
ruled out by cosmic censorship hypothesis \cite{censore}). We will
see later that the spacetime can be extended through the throat
hypersurface in such a way that these singularities are excluded.

The eigenvalues of the Kerr metric tensor (in a optional point with coordinates $(t,r,\theta,\varphi)$) are
\begin{equation} \label{eq2}
\begin{array}{ll}
\lambda_{1}=\frac{r{^2}+a{^2}\cos^{2}\theta}{r^{2}-2mr+a^{2}}
\\
\lambda_{2}=r^{2}+a^{2}\cos^{2}\theta
\\
\lambda_{3}=-\frac{Y(r,\theta)}{r^{2}+a^{2}cos^{2}\theta}
\\
\lambda_{4}=-\frac{Z(r,\theta)}{r^{2}+a^{2}\cos^{2}\theta}
\end{array}
\end{equation}
where
\begin{eqnarray}\label{ll}
\nonumber
Y(r,\theta)&=&\frac{1}{2}((a^{4}(r^{2}-2mr+a^{2})^{2}\sin^{8}\theta
-2a^{2}(a^{2}+a+r^{2})(a^{2}-a +
r^{2})(r^{2}-2mr\\
\nonumber &+&a^{2})\sin^{6}\theta
(a^{8}+(4r^{2}-4)a^{6}+(8mr+1-8r^{2}+6r^{4})a^{4}+(8mr^{3}-4r^{4}\\
\nonumber &+&4r^{6}+8m^{2}r^{2})a^{2}+r^{8})\sin^{4}\theta
+2(a^{2}+a+r^{2})(a^{2}-a+r^{2})(r^{2}-2mr\\
\nonumber
&+&a^{2})\sin^{2}\theta+(r^{2}-2mr+a^{2})^{2})^{1/2}+(a^{4}+(r^{2}-2mr)a^{2})\sin^{4}\theta
+(-a^{4}\\
&+&(-2r^{2}-1)a^{2}-r^{4})sin^{2}\theta+a^{2}-2mr+r^{2}) ,
\end{eqnarray}
and
\begin{eqnarray}
\nonumber
Z(r,\theta)&=&\frac{1}{2}(-(a^{4}(r^{2}-2mr+a^{2})^{2}\sin^{8}\theta
-2a^{2}(a^{2}+a+r^{2})(a^{2}-a+r^{2})(r^{2}-2mr\\
\nonumber &+&a^{2})\sin^{6}\theta+(a^{8}+(4r^{2}-4)a^{6}+(8mr+1-8r^{2}+6r^{4})a^{4}+(8mr^{3}-4r^{4}\\
\nonumber &+& 4r^{6}+8m^{2}r^{2})a^{2}+r^{8})\sin^{4}\theta
+2(a^{2}+a+r^{2})(a^{2}-a+r^{2})(r^{2}-2mr\\
\nonumber &+&a^{2}) \sin^{2}\theta
+(r^{2}-2mr+a^{2})^{2})^{1/2}+(a^{4}+(r^{2}-2mr)a^{2})\sin^{4}\theta
+(-a^{4}\\
&+&(-2r^{2} -1)a^{2}-r^{4})\sin^{2}\theta+a^{2}-2mr+r^{2}),
\end{eqnarray}
while the eigenvalues of the deformed metric (\ref{eq1}) are
\begin{eqnarray}\label{eq3}
\nonumber
\lambda'_{1}&=&\frac{r{^2}+a{^2}\cos^{2}\theta}{r^{2}+\varepsilon^{2}r^{2}+2r\varepsilon-2mr-2mr\varepsilon+a^{2}}
\\
\nonumber \lambda'_{2}&=&r^{2}+a^{2}\cos^{2}\theta
\\
\lambda'_{3}&=&-\frac{Y(r,\theta)}{r^{2}+a^{2}cos^{2}\theta}
\\
\nonumber
\lambda'_{4}&=&-\frac{Z(r,\theta)}{r^{2}+a^{2}\cos^{2}\theta}
\end{eqnarray}

\begin{figure}
\includegraphics[scale=.35]{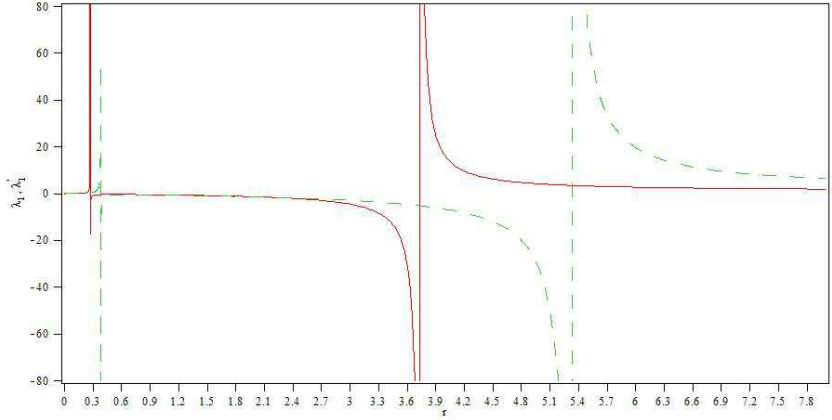}
\caption{$\lambda_{1}$ and $\lambda'_{1}$ are plotted against r
for $\varepsilon=-0.3$, $m=2$ and $a=1$. The solid line denotes
$\lambda_{1}$ and the dashed line corresponds to $\lambda'_{1}.$
}\label{fig1}
\end{figure}

Comparing relations (\ref{eq2}) and (\ref{eq3}) we fined that
\begin{eqnarray}
\nonumber\lambda_{1} \neq \lambda'_{1} \hspace{.1cm} ,
\end{eqnarray}
while
\begin{eqnarray}
\nonumber\lambda_{2}=\lambda'_{2} \hspace{.1cm} ,\\
\nonumber\lambda_{3}=\lambda'_{3} \hspace{.1cm} ,
\end{eqnarray}
 and
\begin{eqnarray}
 \nonumber\lambda_{4}=\lambda'_{4} \hspace{.1cm} .
\end{eqnarray}
The signature of the deformed metric (\ref{eq1}) depends on the
sign of the $\lambda'_{1}$. $\lambda_{1}$, which together with
$\lambda'_{1}$ is shown in Fig. 1 for $m=2$, $a=1$,
$\varepsilon=-0.3$ and $\theta=\frac{\pi}{2}$. It is clear that
the region $r>r_{\varepsilon}^{+}$ has Lorentzian signature and we
have finite redshift at $r=r_{\varepsilon}^{+}$ . The signature of
the metric becomes improper for $r^{+}< r< r_{\varepsilon}^{+}$ ,
which is excluded from the spacetime in the extension proposed
here. Using the famous technique of cut and paste \cite{108p,
109p, 110p}, the spacetime can be extended from
$r=r_{\varepsilon}^{+}$ in such a way the resulting wormhole
structure excludes the intrinsic singularities $r^{\pm}$ and also
the intrinsic ring singularity. Therefore the physical spacetime
includes the region $r>r_{\varepsilon}^{+}$. Also one should note
that the Kretschman scalar ($R_{\mu \nu \lambda \kappa}R^{\mu \nu
\lambda \kappa}$) has a finite value at the throat, implying a
finite gravity at that hypersurface. Since the point $(t,r,\theta,\varphi)$ is a optional point, therefor these conclusion are extendable for whole of spacetime.

\section{Energy-momentum conditions}
Using the Einstein's equations, we readily obtain the
energy-momentum tensor with components:
\begin{eqnarray}
 T_{tt}&=&\frac{1}{8\pi G}G_{tt}=-\frac{1}{8\pi G}\varepsilon
\frac{X(r,\theta)}{(a^{2}-2mr+r^{2})^{2}(a^{2}\cos^{2}\theta +
r^{2})^{4}}\\
\nonumber T_{rr}&=&\frac{1}{8\pi G}G_{rr}\\
&=&\frac{1}{8\pi G}\varepsilon
\frac{Y(r,\theta)}{(r^{2}+a^{2}\cos^{2}\theta)^{2}((1+\varepsilon)^{2}r^{2}-2mr(1+\varepsilon)+a^{2})(a^{2}-2mr+r^{2})}\\
T_{\theta\theta}&=&\frac{1}{8\pi G}G_{\theta\theta}=\frac{1}{8\pi
G}\varepsilon
\frac{Z(r,\theta)}{(a^{2}-2mr+r^{2})^{2}(r^{2}+a^{2}\cos^{2}\theta)^{2}}\\
T_{\varphi\varphi}&=&\frac{1}{8\pi
G}G_{\varphi\varphi}=\frac{1}{8\pi G}\varepsilon
\frac{W(r,\theta)}{(a^{2}-2mr+r^{2})^{2}(a^{2}\cos^{2}\theta
+r^{2})^{4}}\\
T_{t \varphi}&=&T_{\varphi t}=\frac{1}{8\pi
G}G_{t\varphi}=-\frac{1}{8\pi G}ma\varepsilon
\frac{V(r,\theta)}{(a^{2}-2mr+r^{2})^{2}(a^{2}\cos^{2}\theta
+r^{2})^{4}}
\end{eqnarray}
where $X$, $Y$, $Z$, $W$ and $V$ are functions of coordinate $r$
and $\theta$. The eigenvalues of the matrix $T_{\mu \nu}$ are
\begin{eqnarray}
\lambda_{1}&=&\frac{1}{16 \pi G} \varepsilon
\frac{X'(r,\theta)}{(a^{2}-2mr+r^{2})^{2}(a^{2}\cos^{2}\theta
+r^{2})^{4}}\\
\lambda_{2}&=&\lambda_{+}=\frac{1}{8\pi G}\varepsilon
\frac{Y'(r,\theta)}{(r^{2}+a^{2}\cos^{2}\theta)^{2}((1+\varepsilon)^{2}r^{2}-2mr(1+\varepsilon)+a^{2})(a^{2}-2mr+r^{2})}\\
\lambda_{3}&=&\frac{1}{8\pi
G}\varepsilon\frac{Z'(r,\theta)}{(a^{2}-2mr+r^{2})^{2}(r^{2}+a^{2}\cos^{2}\theta)^{2}}\\
\lambda_{4}&=&\lambda_{-}=\frac{1}{16 \pi G} \varepsilon
\frac{W'(r,\theta)}{(a^{2}-2mr+r^{2})^{2}(a^{2}\cos^{2}\theta
+r^{2})^{4}}
\end{eqnarray}
where $X'$, $Y'$, $Z'$ and $W'$ are functions of coordinate $r$
and $\theta$. What can be inferred from this eigenvalues and is
important to us is that first; all the eigenvalues are
proportional to the parameter $\varepsilon$, therefore the
eigenvalues tend to zero if $\varepsilon \rightarrow 0$ (as the
fluid becomes diluted). Second; due to the non-zero spatial
eigenvalues, the distribution can not be a dust. Third; because of
the unequal pressure components, this tensor does not describe a
perfect fluid. Therefore, the proposed spacetime, contains an
imperfect and anisotropic fluid.

Energy-momentum conditions and in particular their status in
wormhole spacetimes are extensively discussed in Morris and Thorne
\cite{MT} and Visser \cite{10magh}. Here, we apply the general
procedure to the deformed metric (\ref{eq1}). Calculating the
eigenvectors of the matrix $T_{\mu \nu}$  we find that the
eigenvector which corresponds to the eigenvalue $\lambda_{-}$ is
timelike. Then the energy density $\rho$ is given by the relation
\begin{equation} \label{eq6}
\rho=\lambda_{-}
\end{equation}
and the components of pressure are
\begin{equation} \label{eq7}
\begin{array}{ll}
p_{1}=\lambda_{2}\\
p_{2}=\lambda_{3}\\
p_{3}=\lambda_{+}
\end{array}
\end{equation}
For any non-spacelike vector $v^{\mu}$, the weak energy condition
reads \cite{10magh, Carroll}
\begin{equation}
T_{\mu\nu}v^{\mu}v^{\nu} \geq 0 \hspace{0.5cm} \Longrightarrow
\hspace{0.5cm} \rho \geq 0 \hspace{0.5cm} and \hspace{0.5cm} \rho
+ p _{i} \geq 0 \hspace{0.2cm} ,\hspace{0.5cm}   i=1, 2, 3
\end{equation}
Using (\ref{eq6}) and (\ref{eq7}), it can be seen that although
$\rho\geq0$, $\rho+p_{2}\geq 0$ and $\rho+p_{3}\geq0$, the weak
(and therefore null and strong) energy conditions are violated in
the whole physical spacetime due to the fact that $\rho+p_{1}<0 $
everywhere in the spacetime.
\section{Summary and conclusion}\label{NLmatchingFFtex}
In this paragraph we summarize the results of our calculation for
the radial deformation of the Kerr metric
($g_{rr}(r)=\frac{\rho^{2}(r,\theta)}{\Delta(r)} \longrightarrow
\frac{\rho^{2}(r,\theta)}{\Delta((1+\varepsilon)r)}$). It is shown
that the resulting spacetime, represents a family of exact
anisotropic rotating fluid solutions. All solutions are singular
on a ring lying in the equatorial plane (like the ring singularity
of Kerr solution), while the deformed metric has one more
singularity. The curvature invariants at large $r$ tend
to zero for the deformed metric (\ref{eq1}), therefor the spacetime is
asymptotically flat.

 We introduced wormhole geometries that arise for certain values of the
parameters (for example $m=2, a=1 \hspace{.1cm} and \hspace{0.1cm}
\varepsilon=-0.1$). This happened because in this range of
parameters, coordinate and intrinsic singularities are pushed
below the physical domain of interest (i.e. the domain with
Lorentzian signature), and the spacetime is extended beyond $ r
> r_{\varepsilon}^{+}$ (In fact, the new metric(\ref{eq1}) does'nt cover whole spacetime covered by the Kerr metric).
We showed that the resulting wormhole can be thought of as being
formed by two distinct but identical asymptotically flat
spacetimes joined at the throat ($r_{\varepsilon}^{+}$) and we
note that the redshift function, $g_{tt}$, is finite at
$r=r_{\varepsilon}^{+}$. Extending the spacetime in this way will
exclude the innear horizon and ring singulary present in the Kerr
spacetime. Also one should note that the Kretschman scalar
($R_{\mu \nu \lambda \kappa}R^{\mu \nu \lambda \kappa}$) has a
finite value at the throat, implying a finite gravity at that
hypersurface.

We discussed the energy conditions for the supporting
energy-momentum tensor and showed that? like most other wormholes,
null, weak and strong energy conditions are violated. The energy
density, as well as $\rho + p_{i} \hspace{.2cm} (i=2,3)$, however,
are positive everywhere. As a prospect for future work, it would
be useful to investigate in more detail the causal structure of
the deformed Kerr spacetime and try to find a deeper description
of the matter source in terms of known physical fields or fluids.

\end{document}